\documentclass[journal]{IEEEtran}

\usepackage[T1]{fontenc}% optional T1 font encoding

\usepackage[none]{hyphenat}
\usepackage[utf8]{inputenc}
\usepackage[english]{babel}
\usepackage{verbatim}
\usepackage{graphicx}
\usepackage{float}
\usepackage{hyperref}
\usepackage{lipsum}
\usepackage{stfloats}
\usepackage{amsmath,amsthm}
\usepackage{amsfonts}
\usepackage{amssymb}
\usepackage{mathtools}
\usepackage{epstopdf}
\usepackage{booktabs,dcolumn,caption}
\usepackage{multirow}
\usepackage{caption}
\usepackage{subcaption}
\usepackage{ragged2e}
\usepackage{cite}
\usepackage{censor}
\usepackage{array}
\newcolumntype{C}[1]{>{\centering\arraybackslash}p{#1}}
\newcolumntype{L}[1]{>{\RaggedRight\arraybackslash}p{#1}}
\usepackage{wasysym}
\newcommand*\diff{\mathop{}\!\mathrm{d}}
\usepackage{mathrsfs}
\usepackage{xcolor}
\usepackage{colortbl}
\usepackage{color,soul}
\usepackage{caption,hypcap}
\usepackage{cleveref}
\usepackage{suffix}

\DeclarePairedDelimiterX\MeijerM[3]{\lparen}{\rparen}%
{#3\delimsize\vert\,\begin{matrix}#1 \\ #2\end{matrix}}
\newcommand\MeijerG[8][]{%
	G^{\,#2,#3}_{#4,#5}\MeijerM[#1]{#6}{#7}{#8}}

\WithSuffix\newcommand\MeijerG*[7]{%
	G^{\,#1,#2}_{#3,#4}\MeijerM*{#5}{#6}{#7}}
\hypersetup{linkcolor=black,citecolor=black}

\newcommand{\bbGamma}{{\mathpalette\makebbGamma\relax}}
\newcommand{\bbK}{{\mathbb{K}}}
\newcommand{\makebbGamma}[2]{%
  \raisebox{\depth}{\scalebox{1}[-1]{$\mathsurround=0pt#1\mathbb{L}$}}%
}

\begin{document}

\title{Gamma-Shadowed Two-Ray with Diffuse Power Composite Fading Model}

\author{Pamela~Njemcevic%\textsuperscript{\orcidicon{0000-0001-5912-2967}}
        ~and~Almir~Maric%\textsuperscript{\orcidicon{0000-0002-3005-3934}}
%        and~Enio~Kaljic%\textsuperscript{\orcidicon{0000-0003-1902-2608}}
%        ,~\IEEEmembership{Member,~IEEE}
        %and~Vladimir~Lipovac,~\IEEEmembership{Member,~IEEE}
%\thanks{Manuscript received XXX; revised XXX and XXX; accepted XXX. This work was in partial supported by XXX. The associate editor coordinating the review of this paper and approving it for publication was XXX (\it{Corresponding author: Almir Maric}).}
\thanks{P. Njemcevic and A. Maric are with the Department
of Telecommunications, Faculty of Electrical Engineering, University of Sarajevo, Sarajevo, Bosnia and Herzegovina.}
%\thanks{V. Lipovac is with the Department of Electrical Engineering and Computing, University of Dubrovnik, Dubrovnik, Croatia.}
\thanks{e-mail: pamela.njemcevic@etf.unsa.ba;}
}

\maketitle

\begin{abstract}
In this paper, a novel gamma-shadowed two-ray with diffuse power (GS-TWDP) composite fading model is proposed. The model is intended for modeling propagation in the emerging wireless networks working at millimeter wave (mmWave) frequencies, and is obtained as a combination of TWDP distribution for description of multipath effects and gamma distribution for modeling variations due to shadowing. After derivation of the exact probability density function (PDF), cumulative distribution function (CDF) and moment generating functions (MGF) expressions are obtained. Proposed model is verified by comparing the analytically obtained results with those measured at 28 GHz and reported in literature. Two upper bound average symbol error probability (ASEP) expressions are then derived for M-ary rectangular quadrature amplitude modulation (RQAM) by employing Chernoff and Chiani approximations of Gaussian Q-function, and are used to investigate relationship between GS-TWDP parameters and system performance. All the results are verified by Monte-Carlo simulation.
\end{abstract}


\IEEEpeerreviewmaketitle

\section{Introduction}
Over the last decade, mmWave bands have attracted large research interest as candidate bands for implementation of the future mobile networks. However, signals propagating in these bands exhibit different characteristics than those propagating at frequencies below 6 GHz. For example, it is empirically shown that at such high frequencies, multipath-caused fast variations of the received signal about its local mean can oftentimes be worse than those in Rayleigh channels~\cite{Mat11, Mav15}, which is phenomena not notably observed at lower frequencies. Consequently, TWDP is proposed as a substitute to traditional Rayleigh or Rician fading models, for modeling multipath effect on signal propagating within the mmWave bands.

Thereat, Rayleigh and Rician fading models assume that the received signal is composed of many weak diffuse components and none or just one strong specular component, respectively. As such, these models can not be used for modeling worse-than-Rayleigh fading conditions. On the contrary, TWDP assumes that the received signal is composed of two strong specular in addition to diffuse components, which makes the model applicable for modeling not just worse-than-Rayleigh, but also Rayleigh and better-than-Rayleigh fading conditions~\cite{Dur02}. Therewithal, it is shown that TWDP provides better fit than  traditional models to empirical data collected at mmWave frequencies, even when multipath-caused variations are less severe than those observed in Rayleigh channel. The aforementioned is confirmed by ray-tracing simulation when modeling vehicular-to-infrastructure (V2I) communication scenario~\cite{Zoc19-1}, and by measurements, when modeling vehicle-to-vehicle (V2V) communication in urban environment at 60 GHz~\cite{Zoc19-2}. Two indoor mmWave measurement campaigns performed in~\cite{Zoc19} revealed that TWDP model is more adequate than traditional models for characterization of indoor mmWave communication, and that it is also the best choice for modeling near-body mmWave channels, both in the front and in the back region~\cite{Mav15}.

Obviously, TWDP model is highly applicable for description of fast amplitude variations in mmWave bands in varieties of propagation scenarios. However, it takes into account only multipath-caused variations of the received signal about the local mean and as such, it assumes that the local mean value is constant. 
%when the transceiver or objects in the environment are not moving significantly. 
Nevertheless, due to shadowing effect, which manifests when the receiver or objects in the environment notably move, local mean value starts to vary too\footnote{In this paper, just like in~\cite{Jer17}, the term shadowing is not necessarily linked "\textit{to the large-scale
fading phenomena also called shadowing, caused by complete
or partial blockage by obstacles many times larger than the
signal wavelength}", but rather to variations of signal's local mean caused by the moving objects or the moving transceiver.}.  Consequently, when estimating overall system performance metrics (such as error rate probability, outage probability, and so on) in mmWave band, slow variations of the signal’s local mean value also should be considered. 
Accordingly, composite fading models - which takes into consideration both, fast multipath-caused variations of a signal about the local mean and slow shadowing-caused variation of the local mean - are the only appropriate models for characterization of propagation in the  dynamic environments. 
%Despite the aforesaid, composite fading models which employs TWDP distributed multipath are, to date, scarcely investigated %However, in the most of the existing literature, evaluation of system performance in TWDP channels is performed by considering fast signal variations over the local mean exclusively.} Only in (e.g. in~\cite{Esp212, Jer16}). 
In these circumstances, it is necessary to identify shadowing model which could be used in conjunction with TWDP for  appropriate description of propagation at mmWave frequencies. 

Thereat, in the majority of literature, shadowing-caused variations are assumed to be lognormally distributed~\cite{Kos05, Abd99}, since lognormal distribution the most accurately describes corresponding measurement results, both in traditional~\cite{Lib92, Wei02}, as well as in mmWave bands~\cite{Rap17, Rap15}.
However, mathematical form of lognormal PDF is not convenient for analytic manipulations, since it hinders calculation of closed-form expressions for most performance metrics of interest~\cite{Abd99}. As a solution, computationally more suitable distributions (such as gamma, inverse-gamma, inverse-Gaussian etc.) have been proposed as a substitute for lognormal PDF. Among them, gamma is the most frequently employed, since it has computationally convenient form and provides similar fit to empirical data as lognormal PDF in many communication scenarios~\cite{Abd99, Abd11}. 
%\ul{Accordingly, due to similarities of lognormal and Gamma PDFs in modeling real data, with later being much superior for analytic computations, Gamma distribution has been widely used for modeling variations caused by shadowing.} 
So, to date, for varieties of multipath distributions, composite fading models such as Rayleigh-Gamma~\cite{Abd98}, Rice-Gamma~\cite{Kos08}, Nakagami-Gamma~\cite{Kos05},  $\kappa \mu$-Gamma~\cite{Sof11} etc., are proposed by employing gamma PDF for modeling shadowing effect. 
%In~\cite{Esp212}, outage probability for inverse gamma (IG) TWDP composite fading model is derived and used to investigate the impact of shadowing severity on overall system performance. However, empirical confirmations of IG distribution suitability for modeling shadowing effect are scarce, since 

Also, for TWDP distributed multipath, the effect of gamma shadowing on overall system performances is considered within the fluctuating two-ray (FTR) model proposed in~\cite{Jer16}. However, the model assumes that shadowing affects only two specular components, while the mean power of diffuse components remains unaffected, which makes FTR inapplicable for modeling propagation scenarios in which all components within the local mean fluctuate. 

However, when scatterers are in the
vicinity of transceivers, beside the specular
components, diffuse components will also travel alongside most of the way~\cite{Jer17}. Consequently, the eventual channel fluctuations would affect all of them  simultaneously~\cite{Jer17}. So, in the described scenario, as the users or the surrounding objects move, human body or any other type of shadowing will affect not only specular but also diffuse components. Since, in addition, diffuse components in mmWave band 
%In addition, diffuse components in mmWave band are more significant than those in conventional bands~\cite{Wan18} and 
can account for up to 40\% of the total received power~\cite{Lec21, Wan18}, there is a clear motivation to investigate the effect of gamma-distributed shadowing which simultaneously affects both, diffuse and specular components, on  overall performance of systems operating in mmWave bands.
%Accordingly, in the \hl{dynamic} TWDP channel, the overall variations affecting received signal should be described by composite fading model, which should encompass both TWDP distributed multipath fading and the appropriately distributed shadowing. 

Considering the aforesaid, in this paper, a novel composite fading model called gamma-shadowed TWDP (GS-TWDP) is proposed by assuming TWDP distributed signal with its local mean distributed following gamma PDF. 
For the proposed model, the appropriate chief probability functions are derived as  simple closed-form expressions and their applicability for derivation of performance metrics is demonstrated by deriving the upper ASEP bounds for M-ary RQAM modulated signal, obtained using Chernoff and Chiani approximations of a Gaussian Q-function. 
Proposed model is also empirically verified by measurements performed at 28 GHz in~\cite{Sam16}, confirming its applicability for modeling overall signal variations at mmWave frequencies. 

The rest of the paper is organized as follows:
In Section II, GS-TWDP fading model is introduced, followed by derivation of its chief probability functions, i.e. PDF, CDF and MGF. Verification of the proposed model is performed is Section III. In Section IV, approximate M-ary RQAM ASEPs of a signal propagating in GS-TWDP fading channel are deduced, while the impact of different model's parameters on ASEP are investigated in Section V. Conclusions are provided in Section VI.

\section{Statistical characterization of  Gamma-shadowed TWDP fading model}
\subsection{TWDP multipath fading model}
TWDP is multipath fading model introduced by Durgin et al.~\cite{Dur02} for modeling Rayleigh, better-than-Rayleigh and worse-than-Rayleigh fading conditions. It assumes that the received complex envelope $r(t)$ is composed of two strong specular components $v_1(t)$ and $v_2(t)$ and many low power diffuse components:
\begin{equation}
\begin{split}
\label{TWDP_envelope}
        r(t) &= v_1(t) + v_2(t) + n(t) \\
        & = V_1\exp{\left(j\Phi_1\right)} + V_2\exp{\left(j\Phi_2\right)} + n(t)
    \end{split}
\end{equation}
Within the model, specular components are assumed to be mutually independent with constant magnitudes $V_1$ and $V_2$ and phases $\Phi_1$ and $\Phi_2$ uniformly distributed in $[0, 2\pi)$, respectively, while diffuse components are treated as a complex zero-mean Gaussian random process $n(t)$ with the average power $2\sigma^2$. Considering the aforesaid, in a frequency-flat, slowly varying TWDP fading channel, the received envelope has the following PDF (which is obtained after the random variable transformation of the equation~\cite[eq. (6)]{Erm16}):
\begin{equation}
    \label{TWDP_PDF}
    \begin{split}
        p{_{r}}(x) & = e^{-K}\sum _{j=0}^{\infty} \frac{2K^j t_j}{(j!)^2}\left(\frac{K+1}{\Omega}\right)^{j+1} \\
        &\times x^{2j+1} \exp{\left(-\frac{x^2(K+1)}{\Omega}\right)},~x>0
    \end{split}
\end{equation}
where $t_j$ is defined as:
\begin{equation}
    \label{t_j}
    t_j = \sum _{k=0}^j \left(\frac{\Gamma }{\Gamma ^2+1}\right)^k \binom{j}{k} \sum _{l=0}^k \binom{k}{l} I_{2 l-k}\left(-\frac{2\Gamma K }{1+\Gamma^2}\right)
\end{equation}
or~\cite[eq. (14)]{Erm16}:
\begin{equation}
    \label{t_j_integralni}
    t_j = \frac{1}{2\pi} \int_0^{2\pi} \exp{\left(-\frac{2\Gamma K \cos{\alpha}}{1+\Gamma^2} \right)}\left(1+\frac{2\Gamma}{1+\Gamma^2} \cos{\alpha}\right)^j \diff{\alpha}
\end{equation}
if expressed in integral form. Thereat, in (\ref{TWDP_PDF}) - (\ref{t_j_integralni}), $I_{\nu}(\cdot)$ is modified Bessel function of a first kind and order $\nu$, $K$ reflects the ratio  between  powers of  specular and diffuse components and is defined as $K = (V_1^2+V_2^2)/2 \sigma^2$ and $\Gamma=V_2/V_1$ is the ratio between magnitudes of two specular components. 
In such conditions, the average power of $r(t)$ given by (\ref{TWDP_envelope}), i.e. its local mean, is equal to $\Omega=V_1^2+V_2^2+2\sigma^2$, and due to initial assumption on $V_1$ and $V_2$ constancy, it is obviously assumed to be constant over the multipath effect's observation interval.~\footnote{It is worthwhile to mentioned that envelope PDF given by (\ref{TWDP_PDF}) is originally derived in terms of the parameter $\Delta=2V_1V_2/(V_1^2+V_2^2)$, used instead of $\Gamma$. 
However, conventional $\Delta$-based parameterization is not in accordance with model’s underlying physical mechanisms, which complicates observation of the impact of the ratio between different signal components
on a system’s performance metrics~\cite{rad} and causes huge error in estimation of TWDP parameters~\cite{rad1}. Therefore, the analysis in this paper is based on physically justified parameter $\Gamma$.}

\subsection{Gamma shadowing model}

However, when the transceiver and/or the surrounding objects significantly move, due to shadowing effect, the local mean $\Omega$ starts to varies about the constant area mean power $P_r$ too. These variations have been traditionally modeled as a lognormal random variable, with its PDF given as:  
\begin{equation}
    \begin{split}        p_\Omega(x)=\frac{1}{x\sqrt{2\pi\sigma_s^2}} \exp{\left(-\frac{\ln^2{\left(x/P_r\right)}}{2\sigma_s^2}\right)},~~~~x>0
    \end{split}
\end{equation}
where $\sigma_s$ is shadowing standard deviation, whose values are usually reported in literature for different propagation conditions. However, analytical form of lognormal distribution is not appropriate for mathematical manipulations and can not be used for calculation of closed-form expressions for error probabilities of differently modulated signals propagating in lognormally distributed shadowing channels. 

Accordingly, for many years, gamma distribution has been used as a substitute PDF, given as:
\begin{equation}
    \begin{split}
        p_\Omega(x)=\frac{1}{\bbGamma[m]}\left(\frac{m}{\Omega_s}\right)^{m} x^{m-1} \exp{\left(-x \frac{m}{\Omega_s}\right)}, x>0
    \end{split}
\end{equation}
where $\bbGamma[\cdot]$ is a gamma function, while $\Omega_s$ and $m$  represent gamma shadowing mean power in the area of interest and shadowing severity, respectively. If these parameters are chosen by matching means
and variances of the lognormal and gamma PDFs, they can be expressed in terms of $\sigma_s$ and $P_r$ as~\cite{Kos05}: 
%In these circumstances, the relationship between parameters of gamma and lognormal PDFs can be determined by matching the mean and the variance of these PDF, as:
\begin{equation}
  m=\frac{1}{\exp{\left(\sigma_s^2\right)}-1}~~~~ \Omega_s=P_r\sqrt{\frac{m+1}{m}}
\end{equation}
%thus providing close approximation of gamma PDF to lognormal, especially in upper tail region of the PDF~\cite{Kos05}.

%Obviously, in a no shadowing case, i.e. as $\sigma_s\to0$, evidently $m_s\to\infty$ and $\Omega_s\to P_r$, as it should be. On the other side, the gamma PDF can be considered to closely approximate the lognormal PDF, especially in upper tail region of the density, for $m_s>0.5$. This restriction limits the upper values of $\sigma_s$. Very often, the shadow spread or 'dB-spread', $\sigma_{dB}$, is used to characterizes shadow severity instead of shadow standard deviation, where $\sigma_{dB}=8.686\sigma_s$. Therefore, the approximation is practically accepted for $\sigma_{dB}<9$dB (i.e. $m_s>0.5$).

For such obtained parameters, it is shown that gamma PDF closely approximate lognormal distribution, especially in upper tail region of the PDF~\cite{Kos05}. However, its analytical form is more favorable for mathematical manipulation than the form of a lognormal PDF, which all make gamma distribution the promising candidate for modeling shadowing-caused variations in conjunction with TWDP multipath fading.  

\subsection{GS-TWDP composite fading model}
Under the above assumptions,  PDF of the GS-TWDP  envelope $r_c$ can be determined by averaging the conditional TWDP PDF with respect to $\Omega$:
\begin{equation}
    \begin{split}
    \label{anvelopa}
        p_{r_c}(x)=\int_0^\infty p_r(x|\omega) p_\Omega(\omega)\diff{\omega}
    \end{split}
\end{equation}
which can be solved using~\cite[eq. (2.3.16 (1))]{Pru86} as:
\begin{equation}
\label{Gamma_TWDP_envelope_PDF}
\begin{split}
    p{_{r_c}}(x) &= \frac{4e^{-K}}{\bbGamma[m]}
    \sum _{j=0}^{\infty} \frac{K^j t_{j}}{(j!)^2}
          \left((K+1)\frac{m}{\Omega_s}\right)^{\frac{m+j+1}{2}} \\ &\times\bbK_{1+j-m}\left[2x\sqrt{(K+1)\frac{m}{\Omega_s}}\right] x^{m+j}
\end{split}
\end{equation}
where $\bbK_{\nu}(\cdot)$ represents modified Bessel function of the second kind and order $\nu$. 

\begin{figure}
   % \centering
    \begin{subfigure}[b,trim={0.15cm, 0.15cm, 0.15cm, 0.95cm},clip]{0.49\textwidth}
        %\centering
        %\captionsetup{justification=centering}
        \includegraphics[trim={0.15cm, 0.10cm, 0.15cm, 0.0cm},clip,width=0.99\linewidth]{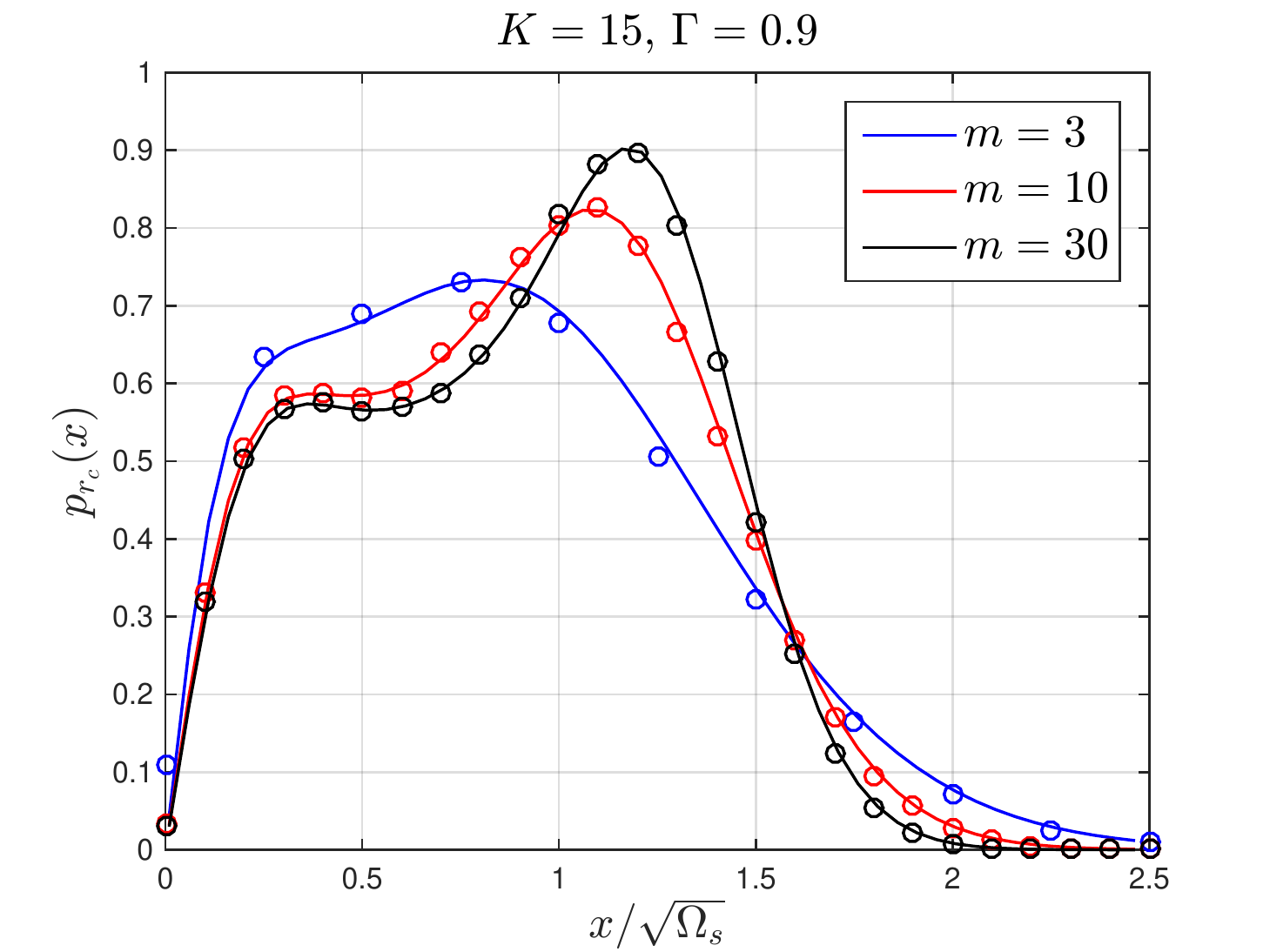}
        \caption{}
        \label{Figure01}
%       \vspace{-12pt}
    \end{subfigure}%%

    \begin{subfigure}[b,trim={0.15cm, 0.1cm, 0.15cm, 0.95cm},clip]{0.49\textwidth}
        %\centering
       % \captionsetup{justification=centering}
        \includegraphics[trim={0.15cm, 0.10cm, 0.15cm, 0.0cm},clip,width=0.99\linewidth]{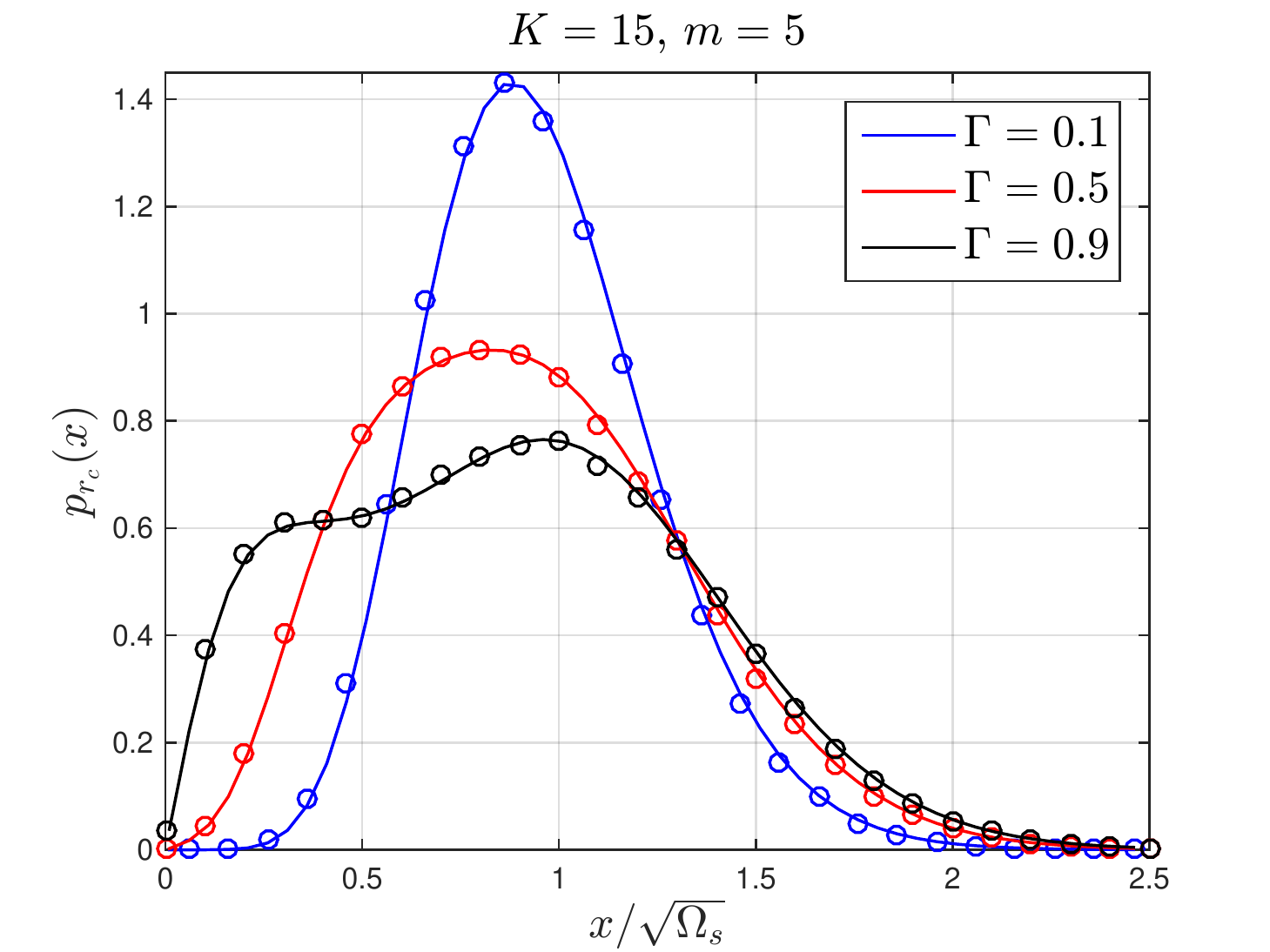}
        \caption{}
        \label{Fig.2a}
       % \vspace{4ex}
    \end{subfigure}%%
%    \vspace{-12pt}
    
    \begin{subfigure}[b,trim={0.15cm, 0.10cm, 0.15cm, 0.95cm},clip]{0.49\textwidth}
     %   \centering
      %  \captionsetup{justification=centering}
        \includegraphics[trim={0.15cm, 0.15cm, 0.15cm, 0.0cm},clip,width=0.99\linewidth]{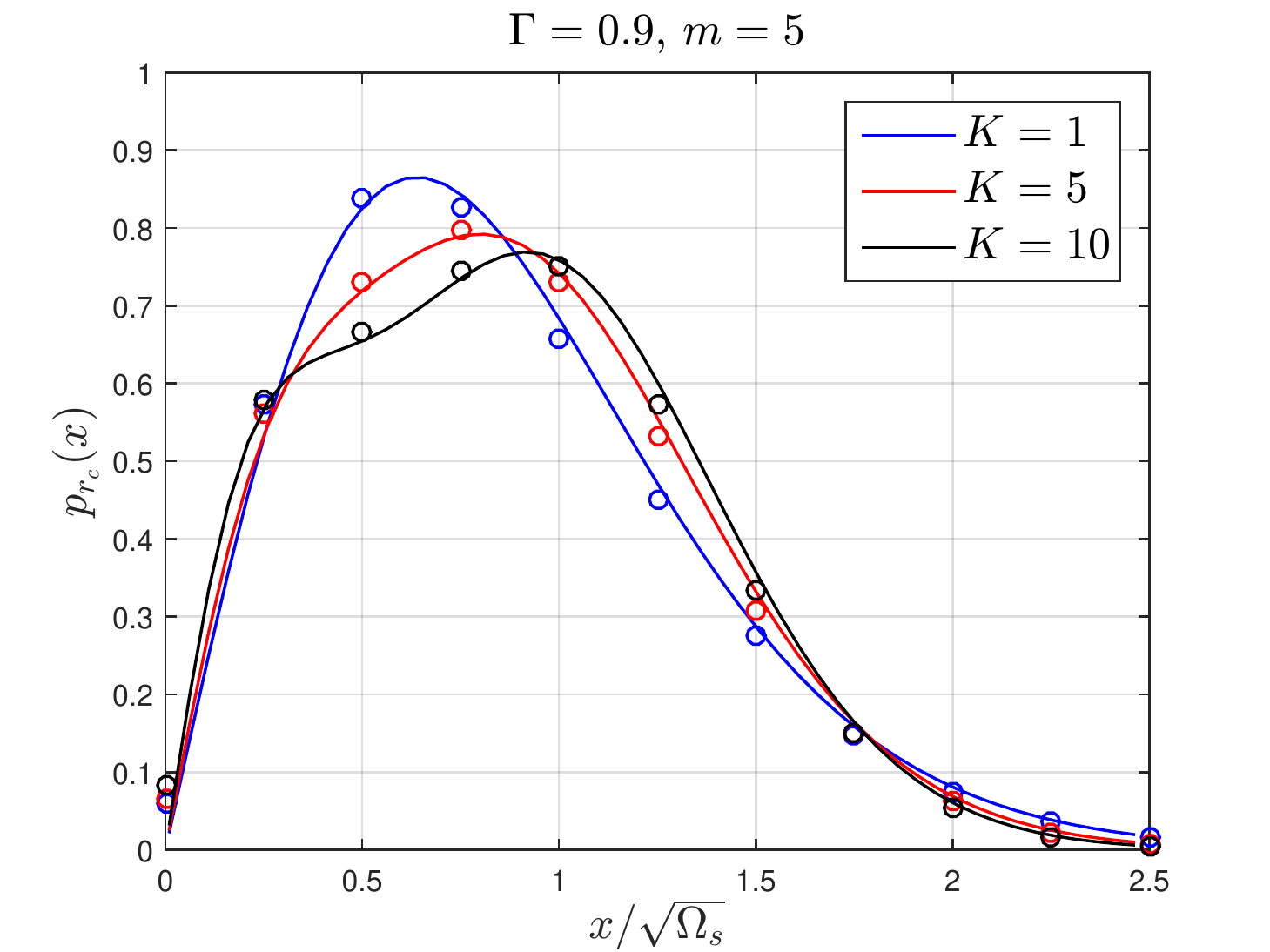}
        \caption{}
        \label{Fig.2b}
        %\vspace{4ex}
    \end{subfigure}
    \caption{Analytical (solid line) and Monte-Carlo simulated (dots) GS-TWDP normalized envelope PDF for different values of: \mbox{(a) $m$ ($K=15$, $\Gamma=0.9$)} \mbox{(b)  $\Gamma$ ($K=15$, $m=5$)} \mbox{(c) $K$ ($\Gamma=0.9$, $m=5$)}} 
\end{figure}

\begin{comment}
\begin{figure}[h]
\includegraphics[trim={0.5cm 0.cm 0.5cm 0.1cm},clip,width=1\linewidth]{PDF1.pdf}
\centering
\caption{} 
\end{figure}

\begin{figure}[h]
\includegraphics[trim={0.5cm 0.cm 0.5cm 0.1cm},clip,width=1\linewidth]{PDF2.pdf}
\centering
\caption{}
\end{figure}

\begin{figure}[h]
\includegraphics[trim={0.5cm 0.cm 0.5cm 0.1cm},clip,width=1\linewidth]{PDF3.pdf}
\centering
\caption{GS-TWDP signal envelope PDF normalized to $\sqrt{\Omega}$, for different values of $K$ and constant $\Gamma$ and $m$, where dots represent Monte-Carlo samples and solid lines are analytical PDFs given (\ref{Gamma_TWDP_envelope_PDF})}
\end{figure}
\end{comment}
Considering the aforesaid, the behaviour of GS-TWDP PDF can be 
observed from Fig. 1, which depicts PDFs expressed by (\ref{Gamma_TWDP_envelope_PDF}) for different arbitrary values
of parameters $K$, $\Gamma$ and $m$ (obtained with maximum 100 summation terms), together with the corresponding simulated PDFs generated by using Monte-Carlo samples which clearly validate derived analytical PDF expression.

Fig. 1 also shows that proposed GS-TWDP distribution exhibits bimodal behavior, which is observed within the empirical PDF obtained by measurements on 28 GHz in~\cite{Sam16}. Accordingly, based on~\cite{Jer16}, it can be served for recreation of the wide heterogeneity of random fluctuations that affect the mmWave radio signal when propagating in the presence of multiple scatterers.

From Fig. 1 also can be observed that bimodality becomes more pronounced with the increment of $K$, $\Gamma$ or $m$, in cases when two other parameters remain unchanged, and that it reaches its maximum when simultaneously $\Gamma \to 1$ and  $K$ is large (which is the combination of corresponding parameters that represents conditions in worse-than-Rayleigh fading channels).    

In addition to the aforesaid, using the identity \cite[eq. 6.561.16]{Gra07}, one can prove that $\int_{0}^{\infty}p_{r_c}(x)\diff{x}=1$. 
It is also interesting to note that derived GS-TWDP envelope PDF expression for $\Gamma = 0$ reduces to a Gamma-shadowed Rician PDF given in~\cite{Kos08}, %since the mean power of the signal simultaneously effected by TWDP small scale fading and Gamma shadowing is equal to $E[\Omega_u] = E[u]\Omega = \Omega$ (due to fact that $\xi$ is  assumed  to  be  a  unit-mean  Gamma  distributed variable) and since $I_{\nu}(0)=1$ for $\nu = 0$ and $I_{\nu}(0)=0$ for all $\nu \neq 0$. 
while for $K=0$, GS-TWDP becomes Gamma-shadowed Rayleigh PDF.
It also can be shown that in no-shadowing case obtained as $m \to \infty$, PDF  (\ref{Gamma_TWDP_envelope_PDF}) reduces to TWDP PDF given in~\cite{Erm16}.

%For ($K=0$), the PDF (\ref{Gamma_TWDP_envelope_PDF}) is identical to K-distribution analyzed in \cite{Abdi00}. On the other side, for ($\Gamma=0$), the PDF (\ref{Gamma_TWDP_envelope_PDF}) reduces to gamma-shadowed Rician PDF \cite{Kos07}. In no shadowing case ($m_s\to\infty$), the PDF (\ref{Gamma_TWDP_envelope_PDF}) is identical to TWDP PDF. 

Finally, it is noteworthy to observe that GS-TWDP and FTR~\cite{Jer17} models are identical only when $m\to\infty$ (which is no shadowing case) and $K\to\infty$ (which is no multipath case), while in all other cases, models will provide different results. 

\subsubsection{Moments}
Derived PDF can be used to obtain $n$-th order moments of the GS-TWDP envelope, by using the expression:
\begin{equation}
    \begin{split}
    \mathbb{E}\{r_c^n\} &=  \int_0^{\infty} x^n p_{r_c}(x)\diff{x}\\
    &=\frac{4e^{-K}}{\bbGamma[m]} \sum_{j=0}^{\infty} \frac{K^j t_{j}}{(j!)^2}\left((K+1)\frac{m}{\Omega_s}\right)^{\frac{m+j+1}{2}}\\ 
    & \times \int_0^\infty \mathbb{K}_{1+j-m}\left[2x\sqrt{(K+1)\frac{m}{\Omega_s}}\right] x^{m+j+n}\diff{x}
    \end{split}
\end{equation}
which can be solved with the help of~\cite[2.16.2(2)]{Pru92} and written in the following form:
\begin{equation}
    \begin{split}
    \label{moments}
    \mathbb{E}\{r_c^n\} &= \frac{e^{-K}}{\bbGamma[m_s]}\sum _{j=0}^{\infty} \frac{K^j t_{j}}{(j!)^2} \left((K+1)\frac{m_s}{\Omega_s}\right)^{-\frac{n}{2}}\\
    &\times\bbGamma\left[1+j+\frac{n}{2}\right] \bbGamma\left[m_s+\frac{n}{2}\right]
    \end{split}
\end{equation}
where $\mathbb{E}\{\cdot\}$ is the expectation operator.
Using (\ref{moments}), it also can be shown that the second moment of the envelope $\mathbb{E}\{r_c^2\}$ is equal to $\Omega_s$. Namely, according to (\ref{moments}), second moment can be written as:
\begin{equation}
    \mathbb{E}\{r_c^2\} = \frac{e^{-K}}{K+1}\sum _{j=0}^{\infty} \frac{K^j t_{j}(j+1)}{(j!)}{\Omega_s}
\end{equation}
where $t_j$ is given by (\ref{t_j_integralni}). After changing the order of integration and summation, the last expression becomes equal to:
\begin{equation}
\label{second_moment}
    \mathbb{E}\{r_c^2\} = \frac{\Omega_s}{K+1}
    \int_0^{2\pi}e^{-a}
    \sum _{j=0}^{\infty} \frac{a^j (j+1)}{(j!)}\diff{\alpha}
\end{equation}
where $a = K\left(1+2\Gamma/(1+\Gamma^2)\right)\cos{\alpha}$. Accordingly, with the help of~\cite[5.2.9(1)]{Pru86}, it can be easily shown that the summation term in (\ref{second_moment}) is equal to $e^a(1+K)$, which makes $\mathbb{E}\{r_c^2\} = \Omega_s$.

\subsubsection{PDF, CDF and MGF}

Based on (\ref{Gamma_TWDP_envelope_PDF}) and the above conclusion, it is now possible to obtaine closed-form expressions for GS-TWDP PDF of the SNR, as well as its closed-form CDF and MGF expressions. 

So, considering the fact that the ratio between instantaneous SNR $\gamma$, and average SNR $\gamma_0$ can be expressed  as:
\begin{equation}
\label{SNR}
    \frac{\gamma}{\gamma_0}=\frac{r_c^2}{\mathbb{E}\{r_c^2\}} =\frac{r_c^2}{\Omega_s}
\end{equation}
PDF of the SNR in GS-TWDP fading channel can be obtained from (\ref{anvelopa}) by simple random variable transformation, as follows:
\begin{equation}
\label{Gamma_TWDP_SNR_PDF}
\begin{split}
    p_{\gamma}(\gamma) &= \frac{2e^{-K}}{\bbGamma[m]}
     \sum _{j=0}^{\infty} \frac{K^j t_{j}}{(j!)^2}
     \left((K+1)\frac{m}{\gamma_0}\right)^{\frac{m+j+1}{2}}  \\ &\times \bbK_{1+j-m}\left[2\sqrt{\gamma(K+1)\frac{m }{\gamma_0}}\right] \gamma^{\frac{m+j-1}{2}}
\end{split}
\end{equation}

On the other side, CDF of the SNR can be derived by integrating (\ref{Gamma_TWDP_SNR_PDF}), using the expression $F_{\gamma}(\gamma)= \int_0^{\gamma} p_{\gamma}(x)\diff{x}$:
\begin{equation}
\begin{split}
    F_{\gamma}(\gamma) &=  \frac{2e^{-K}}{\bbGamma[m]}
     \sum _{j=0}^{\infty} \frac{K^j t_{j}}{(j!)^2}
          \left((K+1)\frac{m}{\gamma_0}\right)^{\frac{m+j+1}{2}} \\ 
          & \times \int_0^{\gamma} \bbK_{1+j-m}\left(2\sqrt{x(K+1)\frac{m}{\gamma_0}} \right) x^{\frac{m+j-1}{2}}\diff{x}
          \end{split}
\end{equation}
which can be solved with the help of~\cite[2.16.3(2)]{Pru92}, as:
\begin{equation}
\begin{split}
\label{Gamma_TWDP_CDF}
F_{\gamma}(\gamma) &=  \frac{e^{-K}}{\Gamma{[m]}}
     \sum _{j=0}^{\infty} \frac{K^j t_{j}}{(j!)^2}\left(\frac{1}{1+j}\left(\frac{(K+1)m}{\gamma_0}\gamma \right)^{1+j}\right.\\
     & \left. \times  _1F_2\left[1 + j; 2 + j, 2 + j - m; \frac{(K+1)m}{\gamma_0} \gamma \right] \right.\\
     & \left. \times \bbGamma[m-j-1] + \frac{1}{m}\left(\frac{(K+1)m}{\gamma_0}\gamma \right)^{m}
     \bbGamma[1+j-m]\right.\\
     & \left. \times _1F_2\left[m; 1 + m, m-j; \frac{(K+1)m}{\gamma_0} \gamma \right]\right)
     \end{split}
\end{equation}
where $_1F_2(\cdot;\cdot,\cdot;\cdot)$ represents generalized hypergeometric function.

Finally, MGF of the instantaneous SNR can be obtained using Laplace transform of (\ref{Gamma_TWDP_SNR_PDF}) (i.e. by solving the integral $\mathcal{M}_{\gamma}(s)= \int_0^{\infty} e^{-s\gamma} p_{\gamma}(\gamma)\diff{\gamma}$), expressed as:
\begin{equation}
    \begin{split}
    \mathcal{M}_{\gamma}(s) &=  
    \frac{2e^{-K}}{\bbGamma{[m]}}
    \sum _{j=0}^{\infty} \frac{K^j t_{j}}{(j!)^2}
    \left((K+1)\frac{m}{\gamma_0}\right)^{\frac{j+m+1}{2}}\\ 
    & \times \int_0^\infty e^{-s\gamma} \mathbb{K}_{1+j-m}\left(2\sqrt{(K+1)\frac{m\gamma}{\gamma_0}}\right) \gamma^{\frac{m+j-1}{2}}\diff{\gamma}
    \end{split}
\end{equation}
which can be solved with the help of~\cite[2.16.8(4)]{Pru92} and~\cite[07.45.27.0003.01, 07.33.17.0007.01]{Wol}, as:
\begin{equation}
    \begin{split}
    \label{MGF}
        \mathcal{M}_{\gamma}(s)&=e^{-K} \sum _{j=0}^{\infty} \frac{K^j t_{j}}{j!}\left(\frac{(K+1)m}{s\gamma_0}\right)^m \\
        & \times U\left[m, m-j,\frac{(K+1)m}{s\gamma_0}\right]
            \end{split}
\end{equation}
where $U[\cdot,\cdot,\cdot]$ is Tricomi confluent hypergeometric function.

It is noteworthy to mention that all derived expressions are given in terms of Bessel and hypergeometric functions, which can be easily evaluated and are efficiently programmed in most standard software packages (e.g., Matlab, Maple and Mathematica)~\cite{Zha18}.

\section{Verification of  GS-TWDP model by  measurement results taken form literature}

After the deviation of the appropriate GS-TWDP statistical expressions, it is also necessary to verify the applicability of proposed distribution for modeling propagation in mmWave bands.
In that sense, GS-TWDP model is fitted to a measurement results performed at 28 GHz and published in~\cite{Sam16}. Obtained results are then expressed in terms of a fitting error calculated using modified version of the Kolmogorov-Smirnov (KS)~\cite{Jer17}, in order to outweigh the fit in amplitude values closer to zero, for which the fading is more severe~\cite{Jer17}.
%are compared to those obtained by fitting FTR model. Thereat, in this paper, the same measurement set reported in~\cite{Sam16}, and the same fitting error metric - modified version of the Kolmogorov-Smirnov (KS) statistic, are used as in~\cite{Jer17}, where FTR model is fitted to a mentioned empirical set. 
%Thereby, in~\cite{Jer17}, a modified version of KS test is proposed in order to outweigh the fit in amplitude values closer to zero, where the fading is more severe. 
Accordingly, goodness of fit between empirical and theoretical CDFs, denoted by $\hat{F}_{r_c}(x)$ and $F_{r_c}(x)$ respectively, is expressed as~\cite{Jer17}:
\begin{equation}
\label{error}
  \epsilon = \max\left|\log_{10}\hat{F}_r(x)-\log_{10}F_{r_c}(x)\right|  
\end{equation}
where $F_{r_c}(x)$ is obtained from (\ref{Gamma_TWDP_CDF}) by using (\ref{SNR}), as:
\begin{equation}
\label{CDF_anvelope}
\begin{split}
    F_{r_c}(x)&=  \frac{e^{-K}}{\Gamma{[m]}}
     \sum _{j=0}^{\infty} \frac{K^j t_{j}}{(j!)^2}\left(\frac{1}{1+j}\left(\frac{(1+K)m}{\Omega}x^2\right)^{1+j}\right.\\
     &\left. \times {}_1F_2\left[1 + j; 2 + j, 2 + j - m; \frac{m(1+K)}{\Omega} x^2\right]\right. \\
     &\left. \times \bbGamma[m-j-1]+ \frac{1}{m}\left(\frac{(1+K)m}{\Omega}x^2\right)^{m}\bbGamma[1+j-m] \right.\\
      &\left. \times  {}_1F_2\left[m; 1 + m, m-j; \frac{m(1+K)}{\Omega} x^2\right]\right)
     \end{split}
\end{equation}

\begin{figure*}[t]
    \centering
    \begin{subfigure}[t]{0.49\textwidth}
    \includegraphics[trim={0.15cm, 0.05cm, 0.15cm, 0.0cm},clip,width=0.99\linewidth]{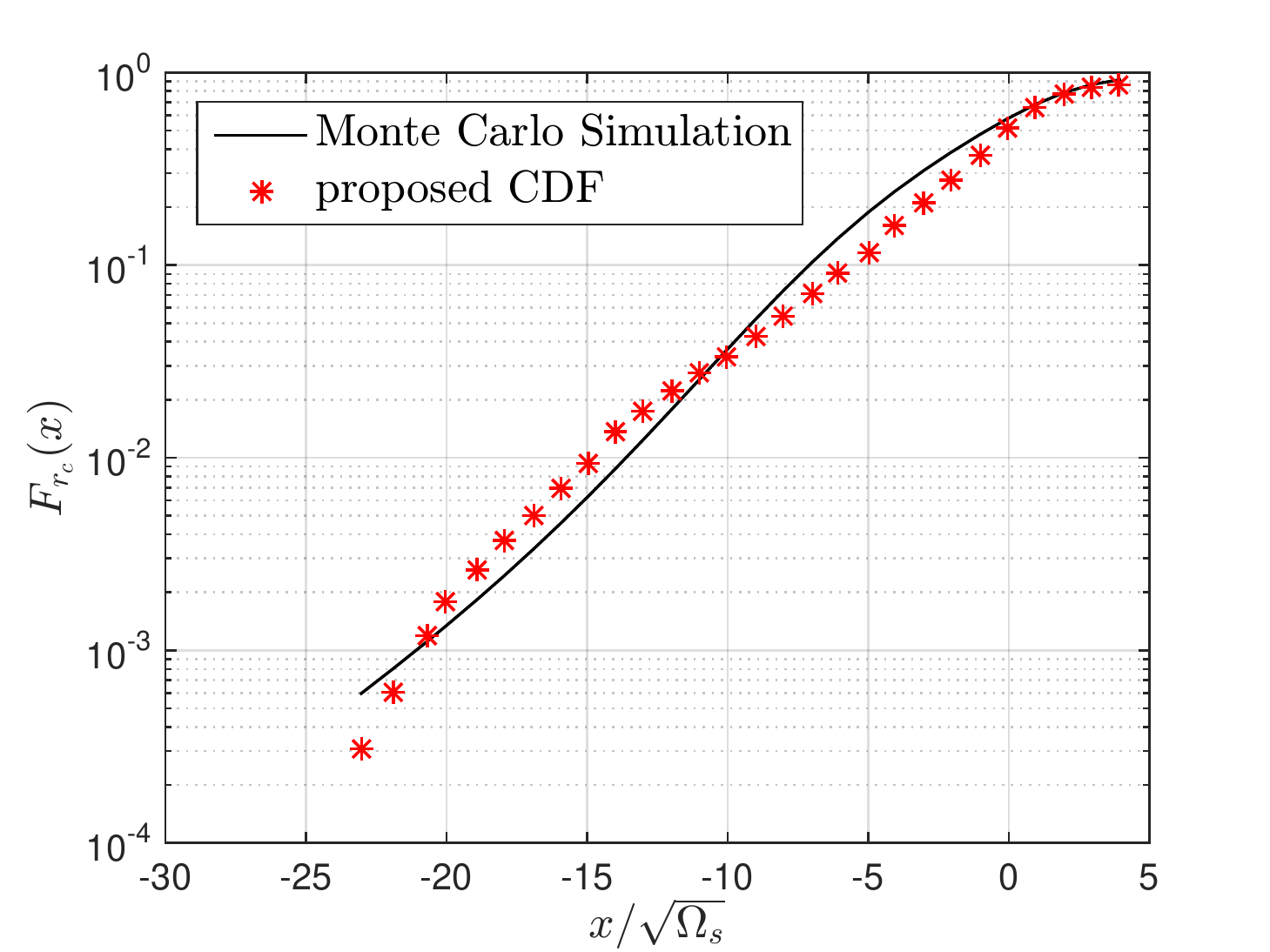}
    \centering
    \caption{}
    %\caption{Best fitting results: $K=15$, $\Gamma = 0.5$, and $m=8$. ($\epsilon=0.285404$)}
    %Empirical~\cite{Sam16} vs theoretical CDFs of the received signal envelope for LOS scenario, for $K=15$, $\Gamma = 0.5$ and $m=8$}
    \end{subfigure}~~~~~
    \begin{subfigure}[t]{0.49\textwidth}
    \includegraphics[trim={0.15cm, 0.05cm, 0.15cm, 0.0cm},clip,width=0.99\linewidth]{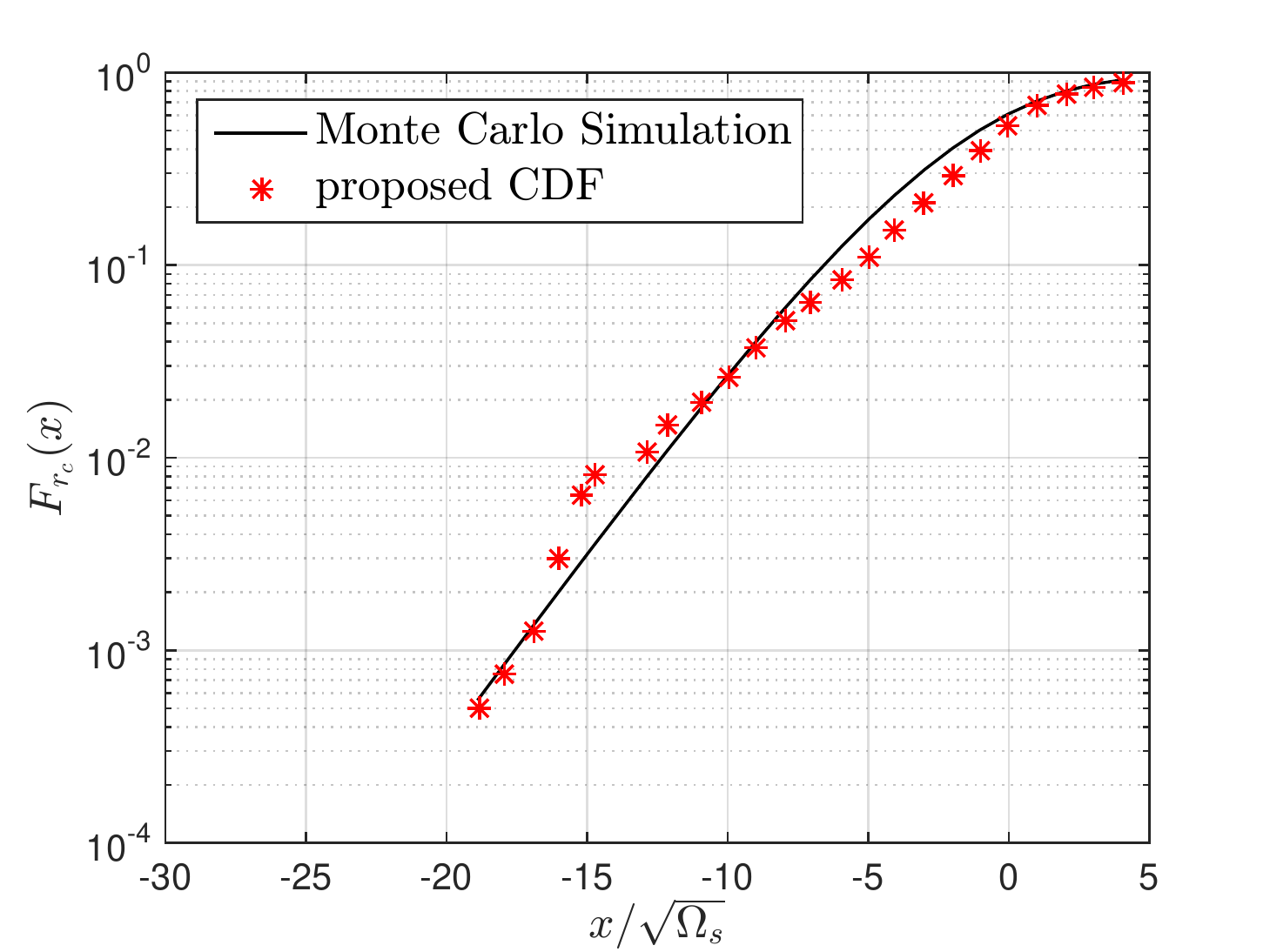}
    \centering
    \caption{}
    %\caption{Best fitting results: $K=22$, $\Gamma = 0.3$, and $m=2$. ($\epsilon=0.280907$)}
    %Empirical~\cite{Sam16} vs theoretical CDFs of the received signal envelope for NLOS scenario, for $K=19$, $\Gamma = 0.5$ and $m=2$}
    \end{subfigure}
    \caption{Fitting of GS-TWDP envelope CDF analytical expression to the empirical data measured in: (a) LOS scenario~\cite[Fig. 6 (LOS)]{Sam16} (b) NLOS scenario~\cite[Fig. 6 (NLOS)]{Sam16}}
    \label{fig:my_label}
\end{figure*}

In Fig. 2, empirical CDFs generated form two different sets of measurements, performed in LOS and NLOS conditions, are compared to the best fitted CDFs generated using (\ref{CDF_anvelope}). Obviously, analytical CDF presented in Fig. 2, obtained for the following set of parameters: $K=15$, $\Gamma = 0.5$ and $m=15$, provides extremely good fit to a given set of LOS measurement results, with the fitting error $\epsilon = 0.22$. 
Similarly, analytically obtained CDF for a set of parameters $K=12$, $\Gamma = 0.15$ and $m=2$ also provides excellent fit to the NLOS measurement results, with the error $\epsilon = 0.24$. 

Besides the aforesaid, it can be noticed that estimated parameters values which provide minimum fitting errors are also in accordance with underlying physical mechanisms. Namely, obtained parameters' values for a LOS case reflect moderate amount of variations caused by multipath, while relatively large value of $m$, as expected in LOS conditions, indicates insignificant shadowing severity. On the country, in the NLOS conditions, parameter $m$ has very small value, which reflects propagation in which shadowing effect is pronounced, as it should be in case of a NLOS propagation.  

\section{Performance analysis}
In order to demonstrate the applicability of obtained GS-TWDP statistical expressions for derivation of error performance metrics, in this section, ASEP of the M-ary RQAM modulation is considered. 
%Derived PDF of the SNR in GS-TWDP fading channel can be used for evaluation of average error rate probabilities, which is demonstrated on the example of the approximate M-ary RQAM ASEP expressions. 
Thereat, M-ary RQAM is chosen as a technique frequently used these days in many wireless communication applications (such as microwave, high-speed mobile communication systems)~\cite{Bil19}, due to its simultaneous power and spectral efficiency. 

Accordingly, for a considered  modulation scheme,
%two different upper bound ASEP expressions are obtained using Chernoff and Chiani approximations of a Gaussian Q-function. 
%Thereat, the aforementioned 
ASEP in GS-TWDP channel can be derived by averaging its conditional SEP in AWGN, $P_e(\gamma)$, over the GS-TWDP PDF given by (\ref{Gamma_TWDP_SNR_PDF}):
\begin{equation}
\label{Pb}
    P_s = \int_0^{\infty } P_e(\gamma)p_{\gamma}(\gamma)\diff{\gamma}
\end{equation}
where $P_e(\gamma)$ can be written as~\cite{Bil19}:
\begin{equation}
\label{ASEP_AWGN}
P_e(\gamma) = 2pQ\left(a\sqrt{\gamma}\right) + 2qQ\left(b\sqrt{\gamma}\right)-4pqQ\left(a\sqrt{\gamma}\right)Q\left(b\sqrt{\gamma}\right)
\end{equation}
and $Q(\cdot)$ is Gaussian Q-function, $p = 1 - 1/M_I$, $q = 1 - 1/M_Q$, $a = \left(6/\left(\left(M_I^2-1\right)+\beta^2\left(M_Q^2-1\right)\right)\right)^{0.5}$, $b=\beta a$ and $\beta$ is the ratio between quadrature and in-phase decision distances, i.e. $\beta = d_Q/d_I$, while modulation order $M$ is equal to $M=M_I\times M_Q$. Following~\cite{Bil19}, (\ref{Pb}) could be calculated as the exact expression given in terms of a bivariate Meijer's G function. However, these functions are very difficult to compute in common software packages. Accordingly, approximations of the involved Q-function could be used 
in order to provide easy-to-compute ASEP upper bound for M-ary RQAM modulation technique. 

In these circumstances, Q-function can be approximated using Chernoff or Chiani approximations:
\begin{equation}
\label{Chernoff}
    Q(x) \approx \frac{1}{2}e^{-\frac{x^2}{2}}
\end{equation}
\begin{equation}
\label{Chiani}
Q(x) \approx \frac{1}{12}e^{-\frac{x^2}{2}} +\frac{1}{4}e^{-\frac{2x^2}{3}} 
\end{equation}
respectively, which, after inserting in (\ref{ASEP_AWGN}), provide the approximate M-ary RQAM conditional SEPs in AWGN is given by~\cite[eq. (12) and (15)]{Bil19}.
These expressions then can be combined with (\ref{Gamma_TWDP_SNR_PDF}) within the expression (\ref{Pb}), and 
after some simple manipulations, expressed in term of the MGF (\ref{MGF}), as:
%the obtained integrals are solved with the help of  ~\cite[2.16.8(4)]{Pru92} and~\cite[07.45.27.0003.01 and 07.45.16.0001.01]{Wol},  providing Chernoff and Chiani approximate M-ary RQAM ASEP expressions for signals propagating in GS-TWDP channel:
\begin{equation}
\begin{split}
    P_{s,{Chernoff}} &\approx p\mathcal{M}_{\gamma}\left(\frac{a^2}{2}\right)
    +q\mathcal{M}_{\gamma}\left(\frac{b^2}{2}\right)\\
    &-pq\mathcal{M}_{\gamma}\left(\frac{a^2+b^2}{2}\right)
\end{split}
\end{equation}
\begin{equation}
\begin{split}
    P_{s,{Chiani}} &\approx \frac{p}{6}\mathcal{M}_{\gamma}\left(\frac{a^2}{2}\right)
    +\frac{q}{6}\mathcal{M}_{\gamma}\left(\frac{b^2}{2}\right)\\
    &-\frac{pq}{6}\mathcal{M}_{\gamma}\left(\frac{a^2+b^2}{2}\right)+\frac{p}{2}\mathcal{M}_{\gamma}\left(\frac{2a^2}{3}\right)\\
    &+\frac{q}{2}\mathcal{M}_{\gamma}\left(\frac{2b^2}{3}\right)-\frac{pq}{2}\mathcal{M}_{\gamma}\left(\frac{2(a^2+b^2)}{3}\right)
\end{split}
\end{equation}
i.e. as a close-form approximate ASEP expressions (\ref{Pb1}) and (\ref{Pb2}).
\begin{figure*}
\begin{equation}
\label{Pb1}
\begin{split}
    P_{s,{Chernoff}} &\approx e^{-K} \sum _{j=0}^{\infty} \frac{K^j t_{j}}{j!}\left(p\left(\frac{K+1}{\gamma_0}\frac{2m}{a^2}\right)^{m}
    U\left[m, m-j,\frac{K+1}{\gamma_0}\frac{2m}{a^2}\right]+q\left(\frac{K+1}{\gamma_0}\frac{2m}{b^2}\right)^{m}\right.\\
    &\left.\times U\left[m, m-j,\frac{K+1}{\gamma_0}\frac{2m}{b^2}\right]
    -pq\left(\frac{K+1}{\gamma_0}\frac{2m}{a^2+b^2}\right)^{m}
    U\left[m, m-j,\frac{K+1}{\gamma_0}\frac{2m}{a^2+b^2}\right]\right)
\end{split}
\end{equation}
\begin{equation}
\begin{split}
\label{Pb2}
    P_{s,{Chiani}} &\approx e^{-K} \sum _{j=0}^{\infty} \frac{K^j t_{j}}{j!}\left(\frac{p}{6}\left(\frac{K+1}{\gamma_0}\frac{2m}{a^2}\right)^{m}
    U\left[m, m-j,\frac{K+1}{\gamma_0}\frac{2m}{a^2}\right]+\frac{q}{6}\left(\frac{K+1}{\gamma_0}\frac{2m}{b^2}\right)^{m}
    U\left[m, m-j,\frac{K+1}{\gamma_0}\frac{2m}{b^2}\right]\right.\\
    &\left.-\frac{pq}{6}\left(\frac{K+1}{\gamma_0}\frac{2m}{a^2+b^2}\right)^{m}
    U\left[m, m-j,\frac{K+1}{\gamma_0}\frac{2m}{a^2+b^2}\right]+\frac{p}{2}\left(\frac{K+1}{\gamma_0}\frac{3m}{2a^2}\right)^{m}
    U\left[m, m-j,\frac{K+1}{\gamma_0}\frac{3m}{2a^2}\right]\right.\\
    &\left.+\frac{q}{2}\left(\frac{K+1}{\gamma_0}\frac{3m}{2b^2}\right)^{m}
    U\left[m, m-j,\frac{K+1}{\gamma_0}\frac{3m}{2b^2}\right]-\frac{pq}{2}\left(\frac{K+1}{\gamma_0}\frac{3m}{2(a^2+b^2)}\right)^{m}
    U\left[m, m-j,\frac{K+1}{\gamma_0}\frac{3m}{2(a^2+b^2)}\right]\right)
\end{split}
\end{equation}
\end{figure*}

%In the limit as $m\to\infty$, i.e. as GS-TWDP becomes TWDP, takes following form:
%\begin{equation}
%\begin{split}
%    &P_{s,{Chernoff}} \approx e^{-K} \sum _{j=0}^{\infty} \frac{K^j t_{j}}{j!} \left(p\left(\frac{K+1}{K+1+\frac{a^2}{2}\gamma_0}\right)^{j+1}\right.\\
%    &\left.+q\left(\frac{K+1}{K+1+\frac{b^2}{2}\gamma_0}\right)^{j+1}\right.\left.-pq\left(\frac{K+1}{K+1+\frac{a^2+b^2}{2}\gamma_0}\right)^{j+1}\right)
%\end{split}
%\end{equation}
\begin{comment}
\begin{equation}
    P_e(\gamma)_{Chernoff} \approx pe^{-\frac{a^2}{2}\gamma} + qe^{-\frac{b^2}{2}\gamma}-pqe^{-\frac{a^2+b^2}{2}\gamma}
\end{equation}
\end{comment}

\section{Numerical results}

In this section, figures showing ASEPs for 4x2 RQAM versus average SNR are presented for different combinations of parameters $K$, $\Gamma$ and $m$, which correspond to a different composite fading channel conditions. Thereby, shown analytical results are obtained by direct evaluation of the expressions (\ref{Pb1}) and (\ref{Pb2}). Additionally, Monte Carlo simulated results are generated using $10^5$ random realizations following GS-TWDP distribution and are presented in order to validate derived analytical expressions.

Accordingly, from Fig. 4 - Fig. 8 can be observed that ASEP expression obtained by using Chiani approximation almost perfectly matches simulated results for all examined combination of GS-TWDP parameters, while derived expression obtained using Charnoff approximation provides the upper ASEP bound for a considered modulation. 

\begin{figure*}[h]
\centering
\begin{subfigure}[t]{0.49\textwidth}
\includegraphics[width=1\textwidth]{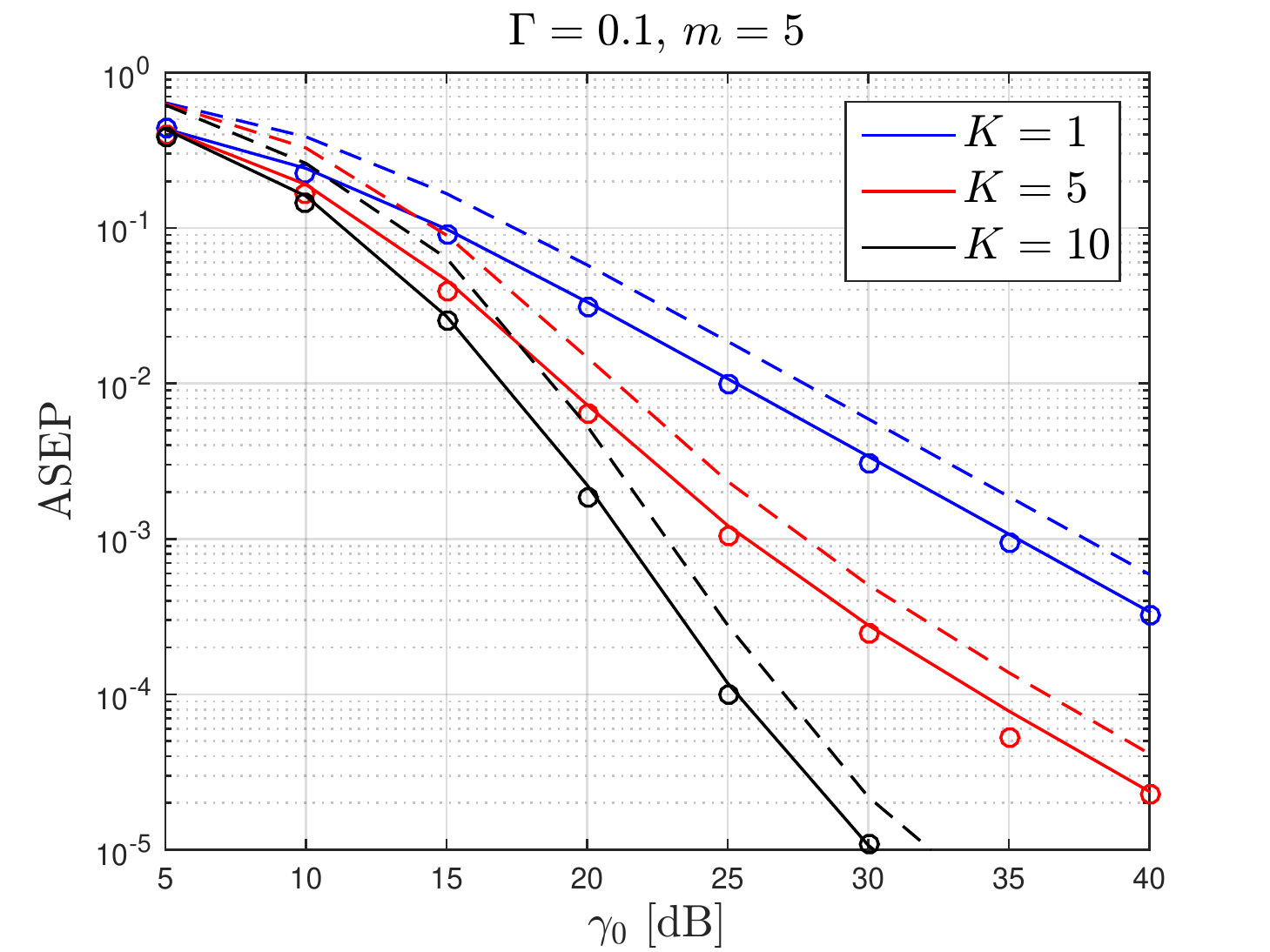}
\centering
\caption{}
%\caption{Chiani approximation (solid line), Chernoff approximation (dotted line), and Monte-Carlo simulation (dots) of 4x2 RQAM ASEP in a GS-TWDP fading channel for different values of $K$ ($\Gamma=0.1, m=5$)}
\end{subfigure}~~~~~
\begin{subfigure}[t]{0.49\textwidth}
\includegraphics[width=1\textwidth]{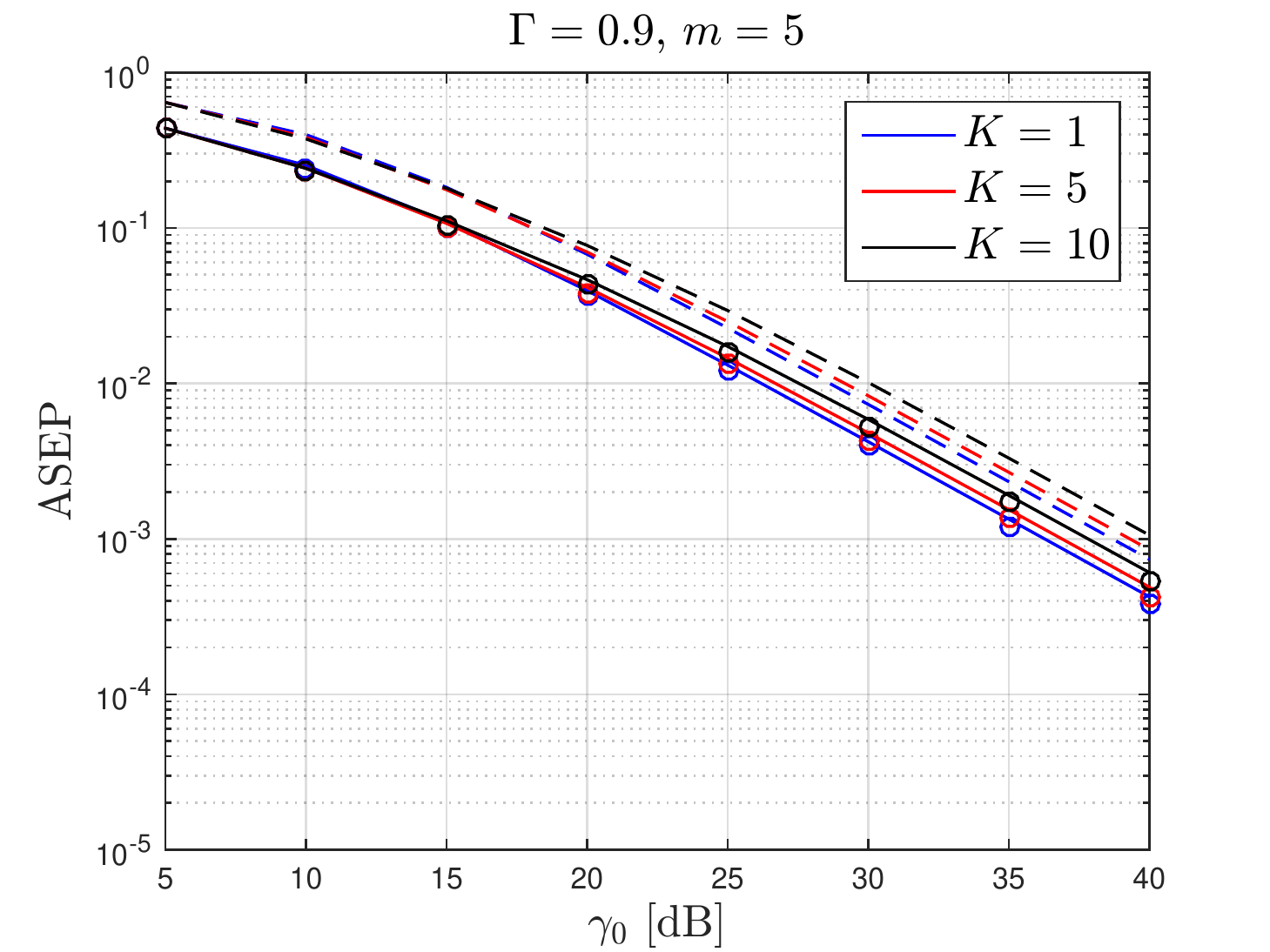}
\caption{}
\centering
\end{subfigure}
\caption{Chiani approximation (solid line), Chernoff approximation (dotted line), and Monte-Carlo simulation (dots) of 4x2 RQAM ASEP in a GS-TWDP fading channel for different values of $K$ and: \mbox{(a) $\Gamma=0.1, m=5$}, \mbox{(b)  $\Gamma=0.9, m=5$}}
\end{figure*}

\begin{figure*}
\begin{subfigure}[t]{0.49\textwidth}
\includegraphics[width=1\textwidth]{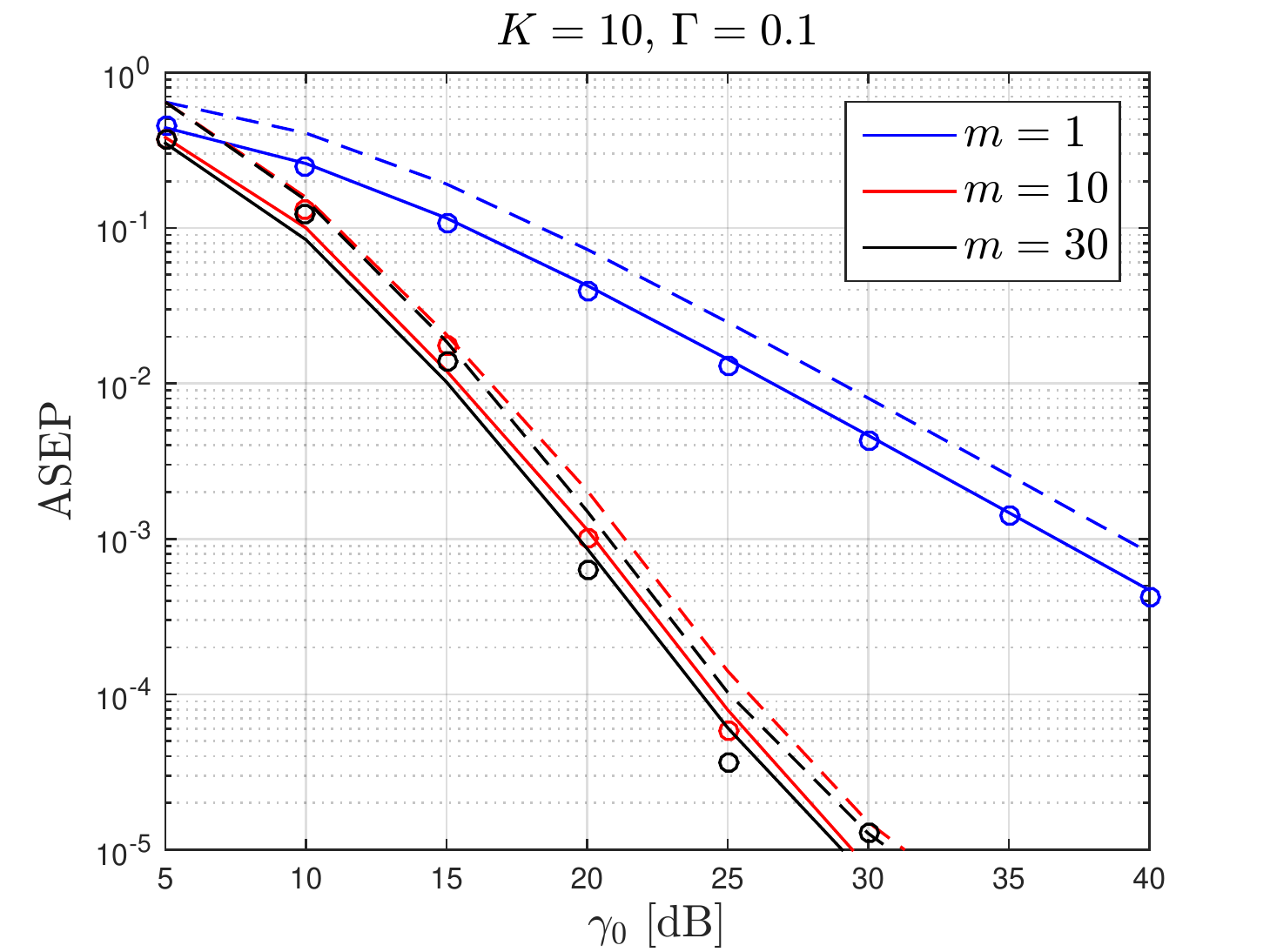}
\centering
\caption{}
%\caption{Chiani approximation (solid line), Chernoff approximation (dotted line) and Monte-Carlo simulation (dots) of ASEP vs. $\gamma_0$ for 4x2 RQAM in a GS-TWDP fading channel, for different values of $m$, constant $K$ and constant small $\Gamma$}
\end{subfigure}~~~~~
\begin{subfigure}[t]{0.49\textwidth}
\includegraphics[width=1\textwidth]{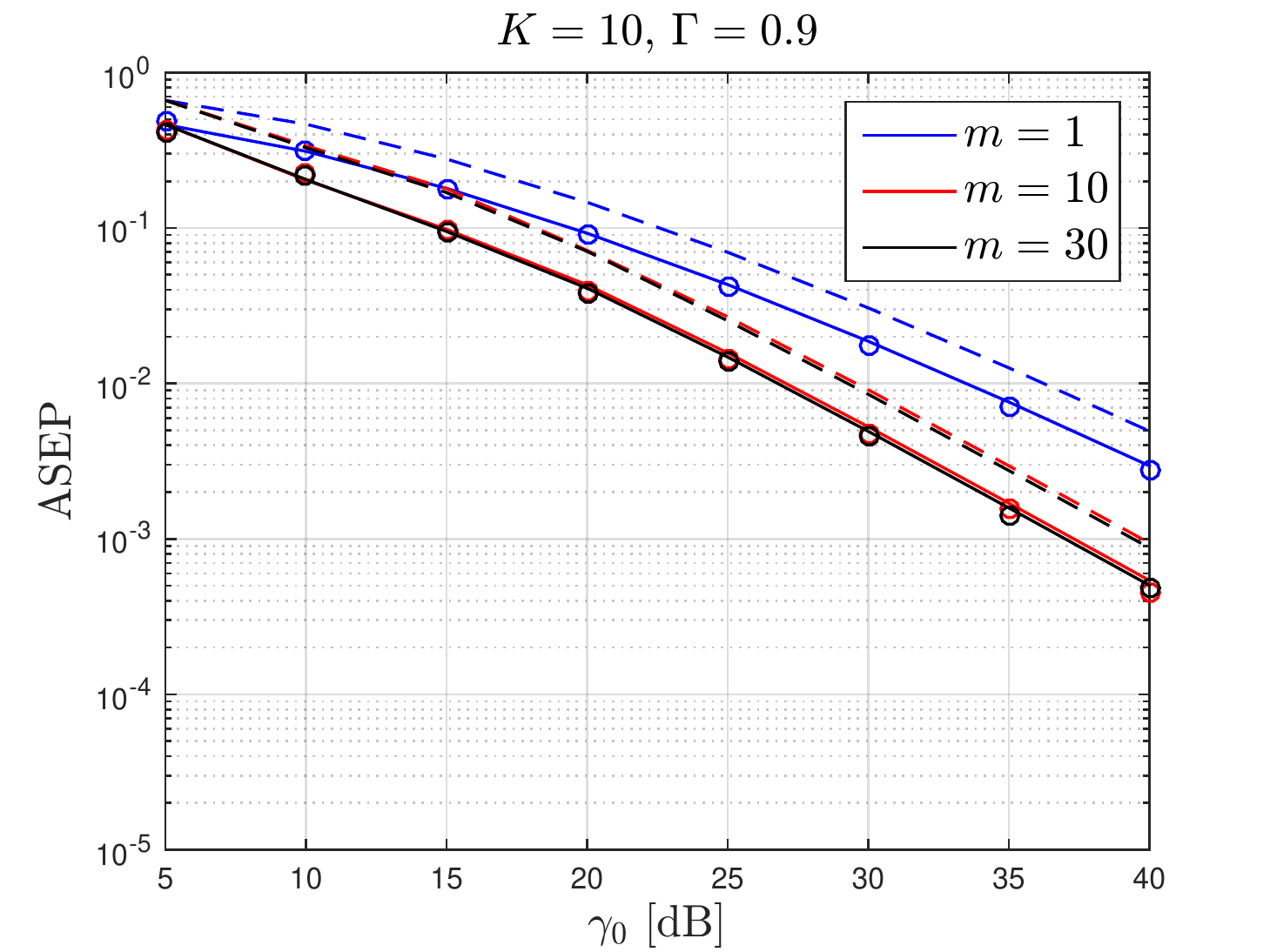}
\caption{}
\centering
\end{subfigure}
\caption{Chiani approximation (solid line), Chernoff approximation (dotted line), and Monte-Carlo simulation (dots) of 4x2 RQAM ASEP in a GS-TWDP fading channel for different values of $m$ and: \mbox{(a) $K = 10$, $\Gamma=0.1$}, \mbox{(b)  $K = 10$, $\Gamma=0.9$}}
\end{figure*}

\begin{figure}
\includegraphics[width=0.49\textwidth]{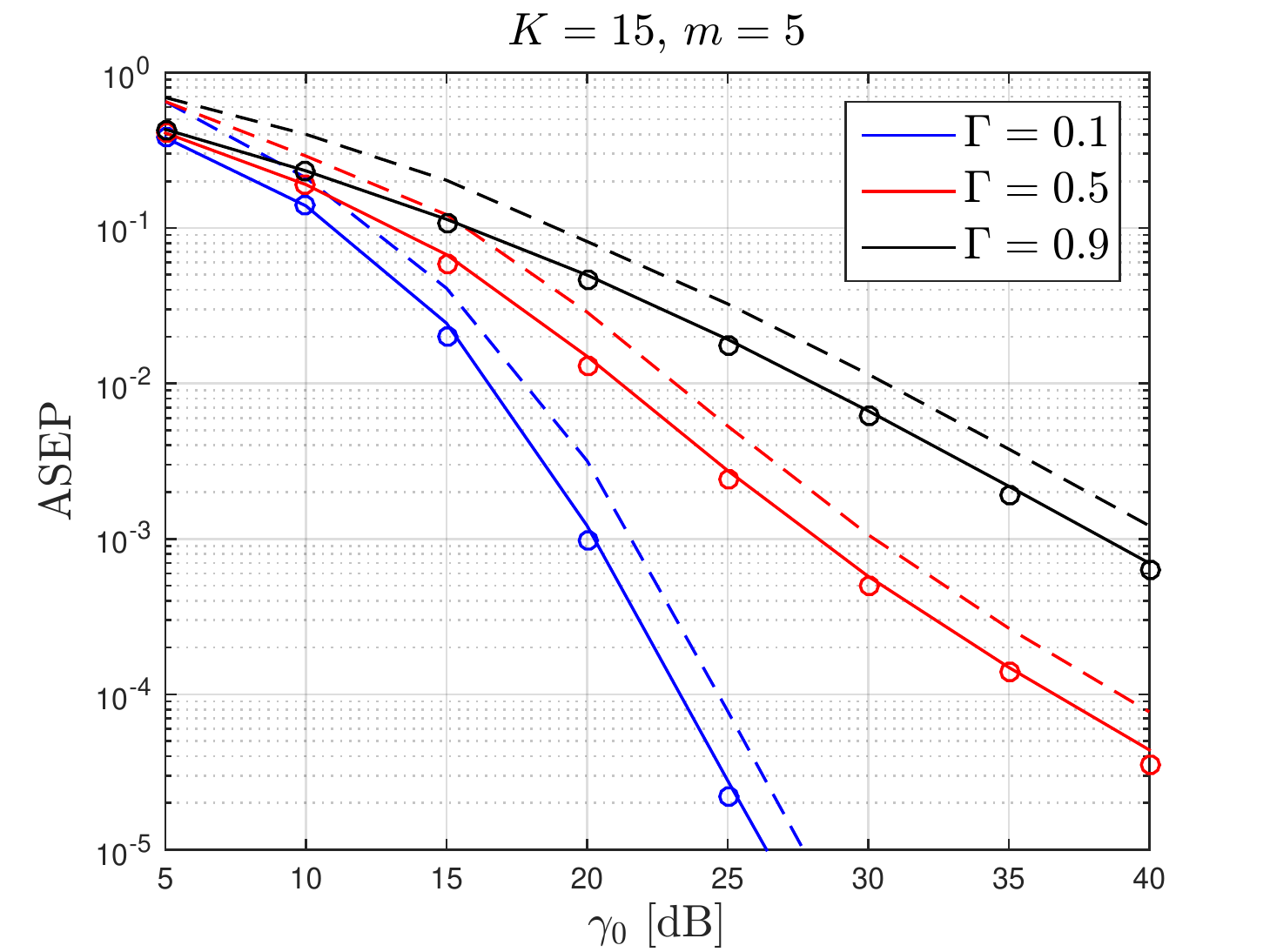}
\centering
%\caption{Chiani approximation (solid line), Chernoff approximation (dotted line) and Monte-Carlo simulation (dots) of ASEP vs. $\gamma_0$ for 4x2 RQAM in a GS-TWDP fading channel, for different values of $\Gamma$ and constant $K$ and $m$}
\caption{Chiani approximation (solid line), Chernoff approximation (dotted line), and Monte-Carlo simulation (dots) of 4x2 RQAM ASEP in a GS-TWDP fading channel for different values of $\Gamma$ ($K=15, m=5$)}
\end{figure}

Thereat, Fig. 4 - Fig. 6 show the impact of multipath severity on system performances, indicating similar behaviour of GS-TWDP to a TWDP model, for a fixed value of parameter $m$. Namely, for small values of $\Gamma$ (which correspond to a light multipath fading severity), increment in $K$ leads to performance improvements. On the contrary, for large $\Gamma$ close to 1 (i.e. when multipath is very severe), performance degrades with the increment of $K$.
In addition, regardless the value of $K$, increment in $\Gamma$ leads to performance degradation.

%The impact of $\Gamma$ increment on ASEP can be observed form Fig. 6, which shows that for small $\Gamma$ values (which occurs in cases when the magnitude of one specular component is significantly larger then the magnitude of the other)  performances are much better then for large values of $\Gamma$, for constant values of $K$ and $m$. 

On the contrary, Fig. 7 and Fig. 8 show the impact of shadowing severity on ASEP. Obviously, when specular and diffuse components experience lighter shadowing-caused fluctuations (expressed by large values of $m$), ASEP improves in respect to cases of shadowing described by small values of $m$.
Thereat, in channels with severe multipath fluctuations (in which $K$ is large and $\Gamma \to 1$), multipath-caused variations dominates and shadowing-caused fluctuations have very small impact on error probability. On the contrary, when multipath-caused variations are less pronounced (i.e. when simultaneously $K$ is large and $\Gamma$ is small), the impact of shadowing on ASEP is much more significant and remarkably worsen performances in respect to those obtained in TWDP channels (for which $m \to \infty$).  
However, regardless the multipath fading severity, when $m$ increases above some value (e.g. $m>10$), shadowing effect on signal can be neglected, and overall performances are going to be pretty much identical to those in TWDP multipath channel.

\section{Conclusion}
In this paper, a novel composite fading model called GS-TWDP is introduced for modeling shadowing- and multipath-caused variation within the emerging mmWave bands. For a proposed model, corresponding PDF, CDF, n-th moment and MGF are derived as a closed-form expressions, given in terms of  functions which can be easily evaluated using standard tools like MATLAB, Mathematica etc. The usability of derived expressions for evaluation of system performance metrics is demonstrated on the example of Chernoff- and Chiani-based M-ary RQAM ASEP upper bound expressions, which are also expressed in terms of  a standardized mathematical function.  
Proposed model is, in addition, empirically verified by utilizing KS test and in literature reported measurement results at 28 GHz, which all make GS-TWDP inevitable candidate for modeling overall signal variations at mmWave frequencies.

\section*{Acknowledgment}
The authors would like to thank Prof. Ivo Kostić for many valuable discussions and advice.

\bibliographystyle{IEEEtran}
\bibliography{Literatura}

%\begin{IEEEbiography}
%	[{\includegraphics[width=1in,height=1.25in,clip,keepaspectratio]{pamela_njemcevic.png}}]{Pamela Njemcevic} graduated from the Faculty of Electrical Engineering, University of Sarajevo, in 2006. She received the M.Sc. and Ph.D. degrees from the Faculty of Electrical Engineering, University of Sarajevo, in 2011 and 2016, respectively. Since 2006, she has been a Teaching Assistant and an Assistant Professor with the Faculty of Electrical Engineering and the Faculty of Transport, Traffic and Communications, University of Sarajevo. She was involved in the implementation of many research projects, related mostly to wireless communications and wireless propagation mechanisms. For two years, she was with EUPM BH, as a Junior EU PPU CARDS Expert for TETRA systems.
%\end{IEEEbiography}

%\begin{IEEEbiography}
%	[{\includegraphics[width=1in,height=1.25in,clip,keepaspectratio]{0_almir_maric_1.jpg}}]{Almir Maric}	received Bachelor's, Master's and PhD degrees in telecommunication engineering from the University of Sarajevo, Bosnia and Herzegovina, in 2010, 2013, and 2020, respectively. He is currently working as Assistant Professor at the Department of Telecommunications at the Faculty of Electrical Engineering, University of Sarajevo. His current research area includes physical channel modeling, communication channel characterization, fading channel and network simulators.
%\end{IEEEbiography}

\end{document}